\documentclass[12pt]{iopart}
\pdfoutput=1

\usepackage{graphicx,url}
\newcommand{\eqref}[1]{(\ref{#1})}
\begin{document}

\title[Neural Spiking and System Identification]{Fitting a recurrent dynamical neural network to neural spiking data: Tackling with the sigmoidal gain function issues}

\author{R. \"{O}zg\"{u}r DORUK}

\address{Department of Electrical and Electronic Engineering, At{\i}l{\i}m University, \.{I}ncek, G\"{o}lba\c{s}\i, Ankara, TURKEY}
\ead{resat.doruk@atilim.edu.tr}
\vspace{10pt}
\begin{indented}
\item[]July 2018
\end{indented}

\begin{abstract}
This is a continuation of a recent study on the modeling of the information coding in sensory system in the brain. The data from a sensory neurons are available as discrete spike timings with no amplitude information. In the simulations, these are generated from a data generator model which has certain differences from the model being estimated. The model under consideration is simpler than the one used as a data generator as it has no sigmoidal gain function parameters. This choice is based on a suggestion from a recent study which states that inclusion of gain functions to the estimation algorithm increases issues such as parameter confounding which leads to performance degradation issues like increased or unpredictable variance changes with different stimuli configurations. To resolve this issue we consider a more generic model that has no sigmoidal gain functions to be estimated. Like that of the first research, we applied a Fourier series stimulus to both data generator and the estimated network. In order to test the performance of the proposed scheme, we have repeated the simulations with different sample sizes, stimulus component counts, amplitude and base frequencies. Mean values of estimation are presented as tables and the statistical analysis results are presented in graphical forms. In addition, since we do not have any true parameter data (for the estimated model) we compare the firing rate responses of both data generator and estimated models to different stimuli. It appeared that the responses of both models are almost the same. 
\end{abstract}

%
 \vspace{2pc}

%
%
%

\section{Introduction}
\subsection{General Neuron Modeling}
\label{sec:gen-neur-modeling}
Neuron modeling is an important milestone in the theoretical or computational neuroscience fields. The nonlinear biological oscillator Hodgkin-Huxley model \cite{hodgkin1952quantitative} is known as a sparking study that triggers the derivation of further simplified or complicated models. Some models only describe the dynamical behaviour of the neurons in consideration. \cite{fitzhugh1961impulses,hindmarsh1984model} models can be considered as an example where the bursting behaviors of a biological neuron is being considered. On the contrary, models including \cite{hodgkin1952quantitative,morris1981voltage,ermentrout2001effects,booth1997compartmental} involve both the dynamical behavior of the membrane potential and some of the internal dynamics due to sodium, potassium and calcium channels and also leakage of chlorine ions.      

In addition there are also some intermediate models such as \cite{rust2005spatiotemporal,hosoya2005dynamic,mante2005independence}. In those, a linear filter and nonlinear cascade is combined. They focus on the computational aspects of a neuron (dynamical features). There are also so called "blackbox models" which are basically a statistical relationship that defines the statistics of the response against a given stimulus. Related examples are \cite{borst1999information,barlow1959possible,fairhall2001efficiency,hassenstein1956systemtheoretische}. 

Another option in neural modeling is the utilization of modified generic neural networks \cite{haykin2004comprehensive}. These models may be either of static feedforward \cite{haykin2004comprehensive,dimattina2011active,dimattina2013adaptive} or recurrent dynamical type \cite{funahashi1993approximation,ledoux2011dynamics,doruk2018}. The generic neural networks are a good option if we are interested in the information coding rather than the biophysical components. In addition, the recurrent neural networks are known to have universal approximation properties \cite{schafer2006recurrent}. Because of that, they are eligible to be used in the modeling of different neurological functions \cite{phdtamw,hancock1997modeling,hancock1999wideband,de2008linking}. 

\subsection{Neural Spiking Issue} 
\label{sec:neur-spik-issue}
Most of the biological neurons transmit information through a phenomenon called as neural spiking \cite{truccolo2005point}. That is a series of action potentials fired by a biological neuron against a stimulus. As repeated action potentials (or bursts) have the same shape, there should be a mechanism that plays a role in  information coding. In sensory neurons, the temporal locations and rates of the spiking (rate of firing of action potentials) are believed to code the transmitted information \cite{0295-5075-22-2-013}. Recent studies also showed that, the neural spiking is not a deterministic process due to stochasticity in ion channels (sodium, potassium, clacium etc.) and synapses \cite{herz2006modeling}. In addition this stochasticity is found to obey an inhomogeneous Poisson distribution \cite{shadlen1994noise} with neural firing rate as its event rate \cite{uteden2008pointprocesses}. There are numerous studies utilizing the findings of that research.  \cite{dimattina2011active,dimattina2013adaptive} develops a static feed-forward network model for the tonotopic auditory cortex in marmoset monkeys. These models are rather computational which are dealing with the information processing side of the neurons. In \cite{doruk2018}, one can see another model where a continuous time recurrent neural network is considered in the same fashion. In that paper, stimulus is modeled as a Phased Cosine Fourier Series (PCFS). The advantage of last paper over works like \cite{dimattina2011active,dimattina2013adaptive} is that, the time dependent features of the model is taken into consideration. In any case, the main output of the models is the neural firing rate which represented by a set of spikes recorded.       

\subsection{About This Work} 
\label{sec:this-work}

The research is to accompany a recent study where the parameters of a continuous time nonlinear dynamical recurrent neural network is fitted from neural spiking data \cite{doruk2018}. However there were certain issues which are not addressed. One major peculiarity is the assumption that the parameters associated with the sigmoidal gain functions are known. In reality, these parameters are not known and should be estimated. However, another recent study \cite{10.7287/peerj.preprints.27015v1} suggests that parameter confounding issues lead to very inefficient estimates. For example in \cite{doruk2018}, the variance of the estimates are decreasing with the increasing sample sizes (or number of iterations). Contrary to that, the work in \cite{10.7287/peerj.preprints.27015v1} reveals that inclusion of the gain parameters into the evaluation leads to unrelated results (i.e. no definite pattern versus the sample size). So the problem should be reconsidered so that one has a useful generic model with no presumed parameters. In addition, if only the firing rate and the associated spike timings are topics of interest one can try a generic continuous time recurrent neural network without any tunable gain function parameters \cite{funahashi1993approximation,haykin2004comprehensive} in the identification procedures. The parameter estimation procedure is similar to that of \cite{doruk2018} and based on joint maximum likelihood estimation \cite{myung2003tutorial}. 

\section{Materials and Methods}
\subsection{Neural Model in Consideration}
\label{sec:model-in-consd} 
Like in \cite{doruk2018}, the model will also be a recurrent dynamical neural network \cite{beer1995dynamics} with excitatory $ (e) $ and inhibitory $ (i) $ components:
\begin{eqnarray}
\label{eq:gen-ctrnn}
\dot{x}_e&=-\beta_e \left[  x_e+w_{ee}g(x_e)-w_{ei}g(x_i)+c_eu \right]  \\
\dot{x}_i&=-\beta_i \left[  x_i+w_{ie}g(x_e)-w_{ii}g(x_i)+c_iu \right] \nonumber 
\end{eqnarray} 
with $ g(x) $ being a generic sigmoid function as shown below:
\begin{equation}\label{eq:sigmoid}
g(x)=\frac{1}{1+e^{-\alpha x}}
\end{equation}     
where $ \alpha $ is a small scaling constant. It can be as small as $ \alpha=0.001 $. In this model setting, $ x_e $ and $ x_i $ are dimensionless variables representing the dynamics of the excitatory and inhibitory components of the model in \eqref{eq:gen-ctrnn} respectively. This is a small difference from \cite{doruk2018} as the state variables have units in mVs. The input $ u $, represents a generic stimulus which can be modeled as a function of time (such as Fourier Series).  

Like in \cite{doruk2018} the output of the model will be the excitatory firing rate $ r_e $. It can be related to the dynamics of the excitatory neuron by a sigmoidal relationship as shown:

\begin{equation}\label{eq:firing-rate-sigmoid}
r_e=\frac{F_e}{1+\exp(-\alpha x_e)}
\end{equation}

The above is similar to \eqref{eq:sigmoid} with a small difference which is the maximum firing rate parameter $ F_e $. 
As a difference from the model setting in \cite{doruk2018}, the firing b The parameters to be estimated are $ \theta=[\beta_e,\beta_i,w_e,w_i,w_{ee},w_{ei},w_{ie},w_{ii},F_e] $ and their definitions are given in \textbf{Table \ref{tab:theta-definition}} below:

\begin{table}
	\centering
	\caption{The definitions of \underline{to be estimated}  parameters. One should note that all of these parameters are greater than zero.}\label{tab:theta-definition}
	\begin{tabular}{ccc}
		\br
		Parameter & Definition & Unit \\
		\mr
		$ \beta_e $ & Excitatory Unit Time Constant & $ s^{-1} $ \\
		\mr 
		$ \beta_i $ & Inhibitory Unit Time Constant & $ s^{-1} $ \\
		\mr
		$ w_{ee} $ & Self Excitation (Autaptic) Coefficient & None \\
		\mr 
		$ w_{ii} $ & Self Inhibition (Autaptic) Coefficient & None \\
		\mr
		$ w_{ei} $ & Inhibitory to Excitatory Synaptic Weight & None \\
		\mr
		$ w_{ie} $ & Excitatory to Inhibitory Synaptic Weight & None \\
		\mr
		$ c_{e} $ & Excitatory Presynaptic Weight  & None \\
		\mr
		$ c_{i} $ & Inhibitory Presynaptic Weight  & None \\
		\mr
		$ F_{e} $ & Maximum Firing Rate of the Excitatory Unit  & $ s^{-1} $ \\
		\br
	\end{tabular}
\end{table}

The estimation procedure will need a data generator that will generate the required neural spiking data sets to be used in computation. The data should represent a realistic neural spiking event set. As this will be a theoretical study, one can achieve this goal by simulating the same or another neuron model and obtain a set of useful spiking data. Obtaining the data from \eqref{eq:gen-ctrnn} with true values of parameters $ \theta $ will not be considered here as a similar situation already exists in \cite{doruk2018}. In addition, the true values of parameters $ \theta $ will actually be not known. Thus, one can generate the neural spiking data from a different model and train the network in \eqref{eq:gen-ctrnn}. The details will be discussed in \textbf{Section \ref{sec:data-gen-model}}.

\subsection{Neural Spiking and Poisson Processes}
\label{sec:neur-spik-poisson-proc} 
\subsubsection{Inhomogeneous Poisson Processes} \label{sec:poisson-theory} 
It is stated in \textbf{Section \ref{sec:neur-spik-issue}} that neural spiking in sensory neurons largely obeys an Inhomogeneous Poisson Distribution \cite{lewis1979simulation} which can be expressed as shown below:
\begin{equation}
\mbox{Prob}\left[N\left(t+\Delta t\right)-N\left(t\right)=k\right]=\frac{e^{-\lambda} \lambda^{k}}{k!}
\label{eq:inhomogeneous-poisson}
\end{equation}
where
\begin{equation}
\lambda=\int_{t}^{t+\Delta t}r_e\left(\tau\right)d\tau
\end{equation}
is the average number of spikes based on the firing rate $ r_e(t) $ of the excitatory unit and $ N(\tau) $ indicates the cumulative total number of spikes up to time $ \tau $, so that  $ N\left(t+\Delta t\right)-N\left(t\right) $ is the number of spikes within the time interval $ {\left[t,t+\Delta t\right)} $. 

Another interpretation is that, the probability of observing $ k $ spikes in the time interval ${\left(t,t+\Delta t\right)}$ will be given by \eqref{eq:inhomogeneous-poisson}. 

Suppose that one has $ K $ spikes $ (t_1,t_2,\ldots,t_K) $ occured in the interval $ (0, T) $ (here $ 0\leq t_1 \leq t_2 \leq \ldots \leq t_K \leq T $ so $ t $ and $ \Delta t $ become $ t=0 $ and $ \Delta t=T $). 

The probability density function for a given spiking train $ (t_1,t_2,\ldots,t_K) $ can be derived from the inhomogeneous Poisson process \cite{uteden2008pointprocesses,brown2002time} as shown in the following:

\begin{equation}
p\left(t_{1},t_{2},\ldots,t_{K}\right)
=\exp\left(-\int_{0}^{T}r_e\left(t,{ u},\theta\right)dt\right)\prod_{k=1}^K r_e\left(t_k,{u},\theta\right)
\label{eq:lkl-enbrown}
\end{equation}

The above formulation describes how likely a particular spike train $(t_1,t_2,\ldots,t_K)$ is generated by an inhomogeneous Poisson process with event rate $ r_e\left(t,{ u},\theta\right) $. It should be noted that this firing rate depends on the parameters $ \theta $ as well as stimulus $ u $. 

The function in \eqref{eq:lkl-enbrown} is derived from Local Bernoulli approximation which will also be  utilized in the neural spike data generator mechanism. The discussion can be found in \cite{uteden2008pointprocesses} and the simulation approach is beriefly presented in \textbf{Section \ref{sec:data-gen-model}}.

\subsubsection{Maximum Likelihood Methods and Parameter Estimation} \label{sub:jmle-theory} 
The maximum-likelihood estimation of the network parameters $ \theta $ in \textbf{Table \ref{tab:theta-definition}} is based on the likelihood function given by \eqref{eq:lkl-enbrown}, 
which takes the individual spike timings into account. 

We know from estimation theory that maximum likelihood estimation is asymptotically efficient, i.e., reaching the Cram\'er-Rao bound when the data sample size becomes larger. 

In order to improve the statistical content of the obtained data, one will need to collect multiple data elicited from multiple stimuli. 

Suppose that we have, $ N_{it} $ independent stimuli that yielded $ N_{it} $ different spike trains and suppose also that  $ m^{th} $ stimulus ($ m=1,\ldots, N_{it} $) elicits a spike train with a total of $ K_m $ spikes in the time window $ [0,T] $, and the spike timings are given by
${S}_m=\left(t_1^{(m)},t_2^{(m)},\ldots,t_{K_m}^{(m)}\right)$. Using \eqref{eq:lkl-enbrown} one can evaluate the likelihood for the spike train $ S_m $ as:

\begin{equation}
p\left(S_m\mid\theta\right)
=\exp\left(-\int_{0}^{T}r_e^{(m)}\!\left(t\right)dt\right)\prod_{k=1}^{K_m} r_e^{(m)}\!\left(t_{k}^{(m)}\right)
\label{eq:pSm}
\end{equation}  

In the above, $ r_e^{(m)} $ is the firing rate response to the $ m^{th} $ stimulus. This function depends implicitly on the model parameters $ \theta $ and on the stimulus. The left-hand side of \eqref{eq:pSm} emphasizes the dependence on network parameters $\theta$, which is convenient for parameter estimation. If the responses to each individual stimulus $ m $ are independent (this will be a reasonable assumption especially when the duration between consecutive stimuli are sufficiently large). Under all those, the joint likelihood function for $ N_{it} $ spike trains can be written as shown below:

\begin{equation}
L\left({S}_{1},S_{2},\ldots,S_{M} \mid \theta \right)
=\prod_{m=1}^{M}p\left(S_m \mid\theta\right)
\label{eq:joint-likelihood-product}
\end{equation}

To ease the computation, one is recommended to evaluate the natural logarithm of \eqref{eq:joint-likelihood-product} and derive the log-likelihood function as:

\begin{equation}
l \left({S}_{1},S_{2},\ldots,S_{M} \mid \theta\right)
=-\sum_{m=1}^{M} \int_{0}^{T}r_e^{(m)}\!\left(t\right)dt 
+\sum_{m=1}^{M}\sum_{k=1}^{K_m}\ln{r_e^{(m)}}\!\left(t_{k}^{(m)}\right)
\label{eq:complete-likelihood-compact}
\end{equation}

The maximization of the above with respect to the parameter vector $ \theta $ will yield the maximum likelihood estimate. That is:

\begin{equation}
\hat\theta_{ML}=\arg\max_{\theta} \left[ l  \left({S}_{1},S_{2},\ldots,S_{M} \mid \theta\right)\right] 
\label{eq:mle-arg-max-log}
\end{equation} 

The above operation can be performed by MATLAB optimization routines such as \texttt{fmincon}. A zero lower bound should be assigned for each parameter in order to ensure that positive valued results are obtained. 

\subsection{Data Generating Model} 
\label{sec:data-gen-model}
In this research, we will generate the neural spiking data by simulating an inhomogeneous Poisson process of which event rate (firing rate) is computed by simulating a different neuron model with predetermined parameters. In this work, we will achieve this goal by simulating the model in \cite{doruk2018} with its true parameters. In this section we will present a summary for the sake of completeness. 

\subsubsection{Data Generator CTRNN}
\label{sec:data-gen-ctrnn}
The data generator neural network is also a continuous time recurrent neural network in the following form:
\begin{eqnarray}
\label{eq:data-gen-ctrnn-model}
\dot{v}_e&=-\beta_e \left[  v_e+w_{ee}g_e(v_e)-w_{ei}g_i(v_i)+c_eu \right]  \\
\dot{v}_i&=-\beta_i \left[  v_i+w_{ie}g_e(v_e)-w_{ii}g_i(v_i)+c_iu \right] \nonumber
\end{eqnarray}  

where the $ v_e $ and $ v_i $ represent dynamics of the excitatory and inhibitory unit respectively. They can be thought like a membrane potential. The critical difference from \eqref{eq:gen-ctrnn} is related to the sigmoidal gain functions $ g_e(v_e) $ and $ g_i(v_i) $. These have a different form than \eqref{eq:sigmoid} as shown below:

\begin{equation}
g_{j}\left(v_{j}\right)=\frac{\Gamma_j}{1+\exp\left( -a_{j}\left(v_{j}-h_j\right)\right) }\label{eq:sigmoid-generator}
\end{equation}
with $ j $ being either $ e $ for excitatory units or $ i $ for inhibitory units. The firing rate from excitatory unit $ r_e $ is evaluated as:

\begin{equation}\label{eq:data-gen-fir-rate}
r_e=g_e(v_e)=\frac{\Gamma_e}{1+\exp(-a_e(v_e-h_e))}
\end{equation}

The model in \eqref{eq:data-gen-ctrnn-model} and \eqref{eq:sigmoid-generator} are simulated with the true values of parameters from \cite{doruk2018} repeated in \textbf{Tables \ref{tab:true-values-data-gen-nn}} and \textbf{\ref{tab:true-values-data-gen-sg}} for convenience. 

\begin{table}
	\centering
	\caption{ 
		The true values of the parameters of the network model in \eqref{eq:data-gen-ctrnn-model}. These are the parameters to be estimated. 
	} \label{tab:true-values-data-gen-nn}
	\begin{tabular}{cc}
		\br 
		Parameter & True value $\left(\theta\right)$ \\
		\mr
		${\beta_e}$ & ${50}$ \\
		\mr
		${\beta_i}$ & ${25}$ \\
		\mr
		${w_e}$ & ${1.0}$ \\
		\mr
		${w_i}$ & ${0.7}$ \\
		\mr
		${w_{ee}}$ & ${1.2}$  \\
		\mr
		${w_{ei}}$ & ${2.0}$ \\
		\mr
		${w_{ie}}$ & ${0.7}$ \\
		\mr 
		${w_{ii}}$ & ${0.4}$ \\
		\br 
	\end{tabular}    
\end{table}

\begin{table}
	\centering
	\caption{The parameters of the sigmoidal gain functions $ g_j(V) $ in \eqref{eq:sigmoid-generator} for the excitatory (e) and inhibitory (i) neurons of the data generator model.}\label{tab:true-values-data-gen-sg}
	\begin{tabular}{cc}
		\br
		Parameter & Value \\ 
		\mr 
		$ \Gamma_e $ & 100 \\ 
		\mr
		$ a_e $ & 0.04 \\
		\mr
		$ h_e $ & 70 \\
		\mr 
		$ \Gamma_i $ & 50 \\ 
		\mr
		$ a_i $ & 0.04 \\
		\mr
		$ h_i $ & 35 \\ 
		\br
	\end{tabular} 
\end{table}

The firing rate $ r_e $ will be solved from the equation $ r_e=g_e(v_e) $. The approach for generation of the spikes are presented in the next section.

\subsubsection{Spike Generation for Data Collection}
\label{sec:spike-generation-for-data}
In order to have a set of neural spiking data that is to be a representative of a collected data from an in vivo experiment we will need to have a reliable algorithm for spike generation. 

In \textbf{Section \ref{sec:neur-spik-poisson-proc}}, we stated that sensory neural spiking largely obeys an Inhomogeneous Poisson Process. So one can implement a Poisson process simulation algorithm with the firing rate $ r_e $ in \eqref{eq:data-gen-fir-rate} being the Poisson event rate. 

There are numerous methodologies to generate the Poisson events given the event rate $ r_e(t) $. These ranging from discrete simulation \cite{uteden2008pointprocesses} to thinning \cite{lewis1979simulation}.

Discrete simulation may be beneficial when one solves the dynamical models by fixed step solvers such as Euler Integration or Runge-Kutta methods. The only disadvantage of this approach is that, it confines the spikes into discrete time bins. However, if one has a sufficiently small discrete time bin such as $ \Delta t=1 $ ms, the statistical distribution of the spikes should approach to that of an Inhomogeneous Poisson Process \cite{uteden2008pointprocesses}.  
Discrete simulation of neural spiking can be summarized as shown below \cite{doruk2018}:

\begin{enumerate}
	\item Given the firing rate of any neuron as $ r(t) $
	\item Find the probability of firing at time $ t_i $ by evaluating $ p_i=r(t_i)\Delta t $ where $ \Delta t $ is the integration interval. It should be as small as 1 ms. 
	\item Compute a random variable by drawing a sample from a distribution which is uniform between 0 and 1. Define this as $ x_{rand}=U[0,1] $ where $ U $ stands for uniform distribution. 
	\item If $ p_i>x_{rand} $ fire a spike at $ t=t_i $, else do nothing.
	\item Collect spikes as $ S=[t_1,\ldots,t_{N_s}] $ where $ N_s $ will be the number of spikes obtained at a single run of simulation. 
\end{enumerate}

\subsection{Stimulus Model}
\label{sec:stimulus-model}
The stimulus in this research will be of Phased Cosine Fourier Series Type as shown below (same as that of \cite{doruk2018}):
\begin{equation}
I=\sum_{n=1}^{N_U}A_n \cos\left(\omega_{n}t+\phi_{n}\right)\label{eq:cosine-stimulus}
\end{equation}
where $ A_n $ is the amplitude, $ \omega_n=2\pi f_0 n $ is the frequency of the $n^{th}$ Fourier component in $ \frac{{rad}}{\sec} $ and $ \phi_n $ is the phase of the component. Here the amplitude $ A_n $ and the base frequency $ f_0 $ (in Hz) are fixed but the phase $ \phi_n $ will be a randomly chosen from a uniform distribution between $ [-\pi,\ \pi] $ radians. The amplitude parameter $ A_n $ is fixed for all mode $ n $ as $ A_n=A_{\max} $.

\begin{figure}
	\centering
	\includegraphics[scale=0.8]{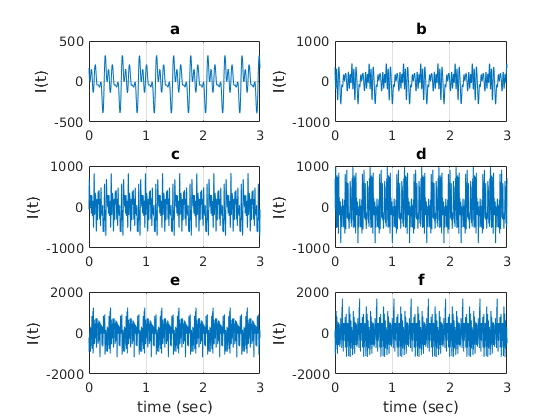}
	\caption{The variation of cosine stimulus for different number of components $ N_U $. The sizes are A) $ N_U=5 $, B) $ N_U=10 $, C) $ N_U=20 $, D) $ N_U=30 $, E) $ N_U=40 $ and F) $ N_U=50 $. The amplitude parameters are $ A_n=100 $, $ f_0=3.333 $ Hz and $ \phi_n $ are randomly assigned from a set uniformly distributed between $ [-\pi,\pi] $.} \label{fig:cosine-stimulus-diff-nu}
\end{figure}

One can see a typical variation of stimulus in \eqref{eq:cosine-stimulus} for different component sizes $ N_U=5\ldots 50 $, $ A_{\max}=100 $ and $ f_0=3.3333 $ Hz in \textbf{Figure \ref{fig:cosine-stimulus-diff-nu}}.

\begin{figure}
	\centering
	\includegraphics[scale=0.8]{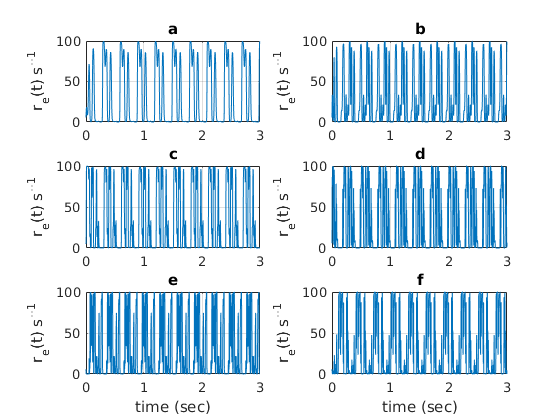}
	\caption{The variation of the firing rate of the excitatory unit against the stimuli shown in \textbf{Figure \ref{fig:cosine-stimulus-diff-nu}}}\label{fig:re-stimulus-diff-nu}
\end{figure}

\begin{figure}
	\centering
	\includegraphics[scale=0.8]{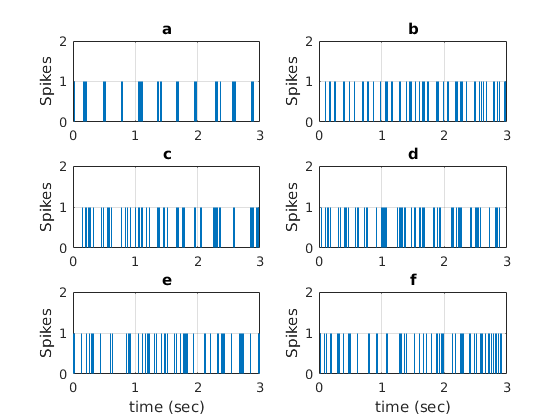}
	\caption{Excitatory neural spiking against the stimuli shown in \textbf{Figure \ref{fig:cosine-stimulus-diff-nu}}. Here the spikes are treated in a binary fashion. Logged as a '1' when a spike exists and '0' when there is no spike.}\label{fig:spikes-stimulus-diff-nu}
\end{figure}

The typical firing rate response of the data generator model in \eqref{eq:data-gen-ctrnn-model}, \eqref{eq:sigmoid-generator} and \eqref{eq:data-gen-fir-rate} with the parameters in \textbf{Table \ref{tab:true-values-data-gen-nn}} and \textbf{\ref{tab:true-values-data-gen-sg}} to the stimuli in \textbf{Figure \ref{fig:cosine-stimulus-diff-nu}} can be seen in \textbf{Figure \ref{fig:re-stimulus-diff-nu}}. In \textbf{Figure \ref{fig:spikes-stimulus-diff-nu}} one can see the neural spiking responses of the excitatory unit against the same stimuli. 

\section{Application}
\subsection{Step-by-step description of the Problem and Simulation}
\label{sub:example-problem-description}        
In this section we will present a step-by-step introduction of the example problem. One can refer to the following on this manner:

\begin{enumerate}
	\item A single run of simulation will last for $ T_f=3 $ seconds. 
	\item The data generator neuron model in \eqref{eq:data-gen-ctrnn-model} and \eqref{eq:sigmoid-generator} will be simulated at the parametric values given in \textbf{Table \ref{tab:true-values-data-gen-nn}} and \textbf{Table \ref{tab:true-values-data-gen-sg}} and firing rate of the excitatory unit is stored as $ r_e^m(t) $ where $ m $ is the current number of simulation. 
	\item Firing rate data $ r_e^m(t) $ is used to generate neural spikes $ S_m $ in the $ m^{th} $ run using the methodology defined in \textbf{Section \ref{sec:spike-generation-for-data}}. This data will be used to compute the likelihood. The number of spikes will be $ K_m $ at the $ m^{th} $ run.  
	\item \label{it:spike-complete} Repeat the simulation $ N_{it} $ times to obtain enough number of spikes. 
	\item The spiking data needed by \eqref{eq:complete-likelihood-compact} will be obtained at the \ref{it:spike-complete}\textsuperscript{th} step. However, the firing rate component of \eqref{eq:complete-likelihood-compact} should be computed at the current iteration of the optimization. 
	\item Run an optimization algorithm on the joint likelihood function \eqref{eq:complete-likelihood-compact} of which objective computes the firing rate at the current iterated value of the parameters but the spikes from \textbf{Step \ref{it:spike-complete}}. This procedure will estimate the parameters of the main model in \eqref{eq:gen-ctrnn} and \eqref{eq:firing-rate-sigmoid} using maximum likelihood estimation as defined in \eqref{eq:mle-arg-max-log}.       
\end{enumerate}

\subsection{Optimization Algorithm}
\label{sub:opt-alg}
Theoretically, any optimization algorithm ranging from gradient descent to derivative free pattern search, simulated annealing or genetic algorithms can be utilized to achieve the computational goals of this research. In the literature and computational software several different implementations are provided to perform the mentioned operations. In this research, we will implement all computations related to optimization and other general operations by MATLAB. Regardless of the type of algorithm, all of the methods converge to a local optimum and thus will require an initial parametric guess. When one has a convex problem all of the initial guesses are expected to converge to the same solution. In the case of problems involving dynamical models, this may or may not be the case. In any case, the main criteria on the choice of the algorithms is the speed of convergence. This is critical regardless of the available computational facilities as we have a huge data to process. The computations are performed using a Hyper Computing Facility (HPC).  Some initial evaluations suggested that local optimizer routines provided by MATLAB's \texttt{fmincon} should be preferred concerning speed and computational resource considerations. MATLAB also provides derivative free optimization techniques such as \texttt{patternsearch} (pattern search technique), \texttt{simulannealbnd} (simulated annealing technique) or \texttt{ga} (genetic algorithm technique) but these may require longer durations to obtain a solution. The \texttt{fmincon} algorithm needed gradient information but it can be provided by itself through finite difference approximations In addition, \texttt{fmincon} and all other optimizers mentioned here are local optimizers and thus multiple initial guesses will be required. When the evaluations are performed on an HPC system, each optimization with a different initial guess can be assigned to a specific processor core. So, a faster parallel computing is achieved and result can be obtained from a single pass. MATLAB Parallel Computing Toolbox will help on this manner. The initial guesses are generated randomly from a uniform distribution. 

\subsection{Simulation Scenarios}
\label{sec:sim-sce}
The data associated with a nominal simulation scenario is presented in \textbf{Table \ref{tab:sim-data}}. 

\begin{table}
	\centering
	\scriptsize
	\caption{Typical data related to the simulation scenario. \textbf{Table \ref{tab:sim-data-varied}}}
	\begin{tabular}{ccc}
		\br 
		Parameter & Symbol & Value \\ 
		\mr
		Simulation Time & $ T_f $ & 3 sec.\\ 
		\mr
		Number of Trials & $ N_{it} $ & 100 \\
		\mr
		\# of Components in Stimulus & $ N_U $ & 5 \\
		\mr
		Method of Optimization & N/A & Interior-Point Gradient Descent (\texttt{fmincon}) \\
		\mr
		\# of Parameters Estimate & Size($ \theta $) & 9 \\
		\mr
		Stimulus Amplitude ($ \mu $A) & $ A_{\max} $ & 100 \\
		\mr 
		Base Frequency & $ f_0 $ & 3.333 Hz \\
		\br
	\end{tabular} 
	\label{tab:sim-data}
\end{table}      

It will be convenient to analyze the influences of change in stimulus component size $ N_U $, amplitude level $ A_{\max} $ and number of samples $ N_{it} $ we will need to repeat the problem for a set of different values of those parameters. The details can be seen in \textbf{Table \ref{tab:sim-data-varied}}.

\begin{table}
	\centering
	\scriptsize
	\caption{The data related to the analysis of the problem for different number of trials $ N_{it} $, number of stimulus components $ N_U $, stimulus amplitude $ A_{\max} $}
	\begin{tabular}{ccc}
		\br 
		Parameter & Symbol & Value(s) \\ 
		\mr
		Number of Trials & $ N_{it} $ & 25, 50, 100, 200, 400 \\
		\mr
		\# of Components in Stimulus & $ N_U $ & 5, 10, 20 \\
		\mr
		Stimulus Amplitude ($ \mu $A) & $ A_{\max} $ & 25, 50, 100, 200 \\
		\mr
		Base Frequency (Hz) & $ f_0 $ & $ \frac{1}{3} $, 1, $ \frac{7}{3} $, $ \frac{10}{3} $, 5 \\
		\br
	\end{tabular} 
	\label{tab:sim-data-varied}
\end{table}

The initial conditions of the excitatory and inhibitory units are not accurately known thus we will assume them as zero i.e. $ x_e(0)=0 $ and $ x_i(0)=0 $. We will repeat the simulation 10 times for each case, so that we will have statistically sufficient number of results for analyses. 

\section{Results}
\label{sec:results-presse}
We will present the numerical results of the maximum likelihood estimation of the parameters of our neuron model in \eqref{eq:gen-ctrnn} in \textbf{Table \ref{tab:theta-definition}} by maximizing the joint likelihood function in \eqref{eq:complete-likelihood-compact}. The optimization is performed using the gradient based interior-point method provided by MATLAB's \texttt{fmincon} algorithm. All the cases in \textbf{Table \ref{tab:sim-data-varied}} are examined under the conditions in \textbf{Table \ref{tab:sim-data}}. The results will be presented as:

\begin{enumerate}
	\item The mean estimated values of $ \theta $ against varying sample size $ N_{it} $, amplitude level $ A_{\max} $, stimulus component size $ N_U $ and base frequency $ f_0 $.  
	\item The standard deviations of the estimated parameters against varying sample size $ N_{it} $, amplitude level $ A_{\max} $, stimulus component size $ N_U $ and base frequency $ f_0 $.
	\item The variations of the mean square error between excitatory firing rate $ r_e $ obtained from data generator model and estimated model. This is required to verify the estimation's accuracy as we do not have any true parameters for the estimated model.     
\end{enumerate}

\subsection{Mean Estimated Values}
\label{sec:mean-estim-values}
In this section, we will present the estimated mean values of all parameters in \textbf{Table \ref{tab:theta-definition}} against varying scenario parameters in \textbf{Table \ref{tab:sim-data}} and \textbf{\ref{tab:sim-data-varied}}. The results are presented in forms of tables starting from \textbf{Table \ref{tab:mean-vals-vs-nit}}. Explanations are given in the table definitions. 

\begin{table}
	\centering 
	\caption{The variation of the estimated parameters against changing sample size $ N_{it} $. The settings of stimulus component size $ N_U $, maximum amplitude parameter $ A_{\max} $ and  base frequency $ f_0 $ are as given in \textbf{Table \ref{tab:sim-data}}.}\label{tab:mean-vals-vs-nit}
	\begin{tabular}{cccccccccc}
		\br
		$ N_{it} $ & $ \beta_e $ & $ \beta_i $ & $ w_e $ & $ w_i $ & $ w_{ee} $ & $ w_{ei} $ & $ w_{ie} $ & $ w_{ii} $ & $ F_e $\\
		\mr
		25 & 38.26 & 25.48 & 53.98 & 20.48 & 7725.53 & 13883.68 & 2796.59 & 4171.99 & 98.58 \\ 
		50 & 36.23 & 25.94 & 55.60 & 20.64 & 7640.84 & 13923.25 & 2637.48 & 4005.38 & 98.87 \\ 
		100 & 36.28 & 26.04 & 55.97 & 20.63 & 7427.70 & 13666.09 & 2533.39 & 3844.03 & 98.76 \\ 
		200 & 36.17 & 26.42 & 55.83 & 20.97 & 7539.17 & 13669.48 & 2555.44 & 3859.42 & 98.47 \\ 
		400 & 36.23 & 26.42 & 55.68 & 20.77 & 7252.09 & 13256.03 & 2428.22 & 3615.60 & 98.72 \\ 
		\br
	\end{tabular}
\end{table} 

\begin{table}
	\centering 
	\caption{The variation of the estimated parameters against changing stimulus component size $ N_U $. The settings of sample size $ N_{it} $, maximum amplitude parameter $ A_{\max} $ and  base frequency $ f_0 $ are as given in \textbf{Table \ref{tab:sim-data}}.}\label{tab:mean-vals-vs-nu}
	\begin{tabular}{cccccccccc}
		\br
		$ N_{U} $ & $ \beta_e $ & $ \beta_i $ & $ w_e $ & $ w_i $ & $ w_{ee} $ & $ w_{ei} $ & $ w_{ie} $ & $ w_{ii} $ & $ F_e $\\
		\mr
		5 & 36.28 & 26.04 & 55.97 & 20.63 & 7427.70 & 13666.09 & 2533.39 & 3844.03 & 98.76 \\ 
		10 & 37.64 & 26.61 & 53.21 & 18.88 & 6858.12 & 12768.65 & 2254.27 & 3339.32 & 99.39 \\ 
		20 & 37.81 & 26.76 & 53.29 & 19.56 & 7131.97 & 13046.52 & 2087.44 & 3300.90 & 99.01 \\ 
		\br
	\end{tabular}
\end{table} 

\begin{table}
	\centering 
	\caption{The variation of the estimated parameters against changing maximum amplitude $ A_{\max} $. The settings of sample size $ N_{it} $, stimulus component size $ N_{U} $ and  base frequency $ f_0 $ are as given in \textbf{Table \ref{tab:sim-data}}.}\label{tab:mean-vals-vs-amax}
	\begin{tabular}{cccccccccc}
		\br
		$ A_{\max} $ & $ \beta_e $ & $ \beta_i $ & $ w_e $ & $ w_i $ & $ w_{ee} $ & $ w_{ei} $ & $ w_{ie} $ & $ w_{ii} $ & $ F_e $ \\
		\mr
		25 & 53.21 & 515.45 & 54.44 & 27.31 & 20896.08 & 38946.13 & 18011.48 & 33635.76 & 51.18 \\ 
		50 & 35.00 & 22.94 & 58.42 & 25.15 & 9653.59 & 18991.63 & 4407.25 & 7983.17 & 95.89 \\ 
		100 & 36.28 & 26.04 & 55.97 & 20.63 & 7427.70 & 13666.09 & 2533.39 & 3844.03 & 98.76 \\ 
		200 & 40.55 & 28.73 & 49.75 & 14.67 & 6537.35 & 11855.27 & 1528.09 & 2215.13 & 99.15 \\ 
		\br
	\end{tabular}
\end{table}

\begin{table}
	\centering 
	\caption{The variation of the estimated parameters against changing base frequency $ f_{0} $. The settings of sample size $ N_{it} $, stimulus component size $ N_{U} $ and  maximum amplitude $ A_{\max} $ are as given in \textbf{Table \ref{tab:sim-data}}.}\label{tab:mean-vals-vs-f0}
	\begin{tabular}{cccccccccc}
		\br
		$ f_{0} $ & $ \beta_e $ & $ \beta_i $ & $ w_e $ & $ w_i $ & $ w_{ee} $ & $ w_{ei} $ & $ w_{ie} $ & $ w_{ii} $ & $ F_e $ \\
		\mr
		$ \frac{1}{3} $ & 30.84 & 26.69 & 69.63 & 31.65 & 8864.84 & 15372.76 & 3447.75 & 5210.14 & 98.58 \\
		$ 1 $ & 34.69 & 27.65 & 59.15 & 23.10 & 7217.08 & 12867.85 & 2378.26 & 3385.87 & 98.97 \\
		$ \frac{7}{3} $ & 36.48 & 26.71 & 56.04 & 19.90 & 6310.13 & 11976.28 & 1901.61 & 2735.95 & 98.46 \\
		$ \frac{10}{3} $ & 36.28 & 26.04 & 55.97 & 20.63 & 7427.70 & 13666.09 & 2533.39 & 3844.03 & 98.76 \\
		$ 5 $ & 38.36 & 23.57 & 52.57 & 18.71 & 7088.79 & 14886.53 & 2462.94 & 4750.91 & 99.14 \\
		\br
	\end{tabular}
\end{table}

\subsection{Change in the Standard Deviation of the Estimates}
\label{sec:std-changes}
\begin{figure}
	\centering
	\includegraphics[scale=0.85]{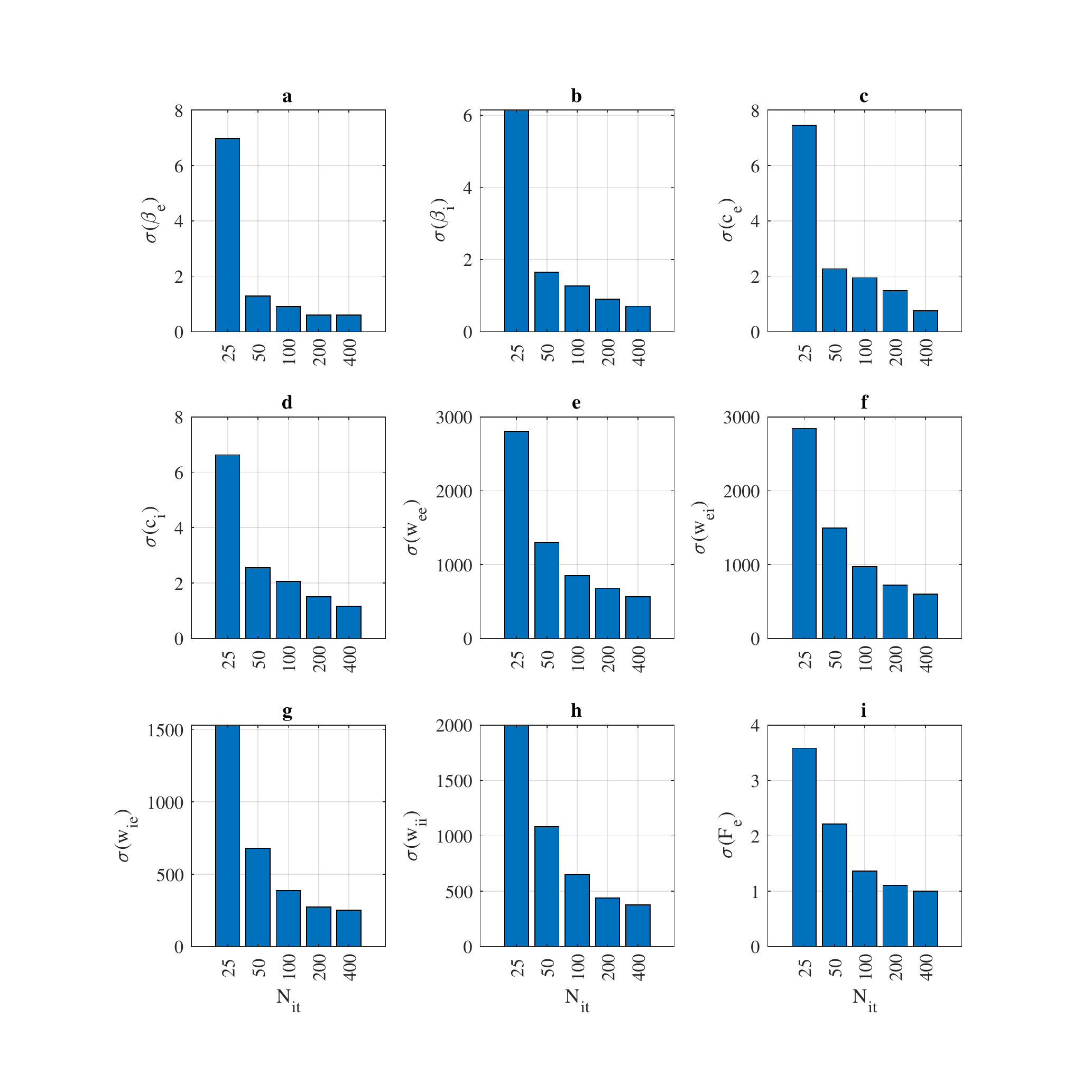}
	\caption{The variation of the standard deviations of the estimated parameters against increasing sample size $ N_{it} $. The settings of stimulus component size $ N_U $, maximum amplitude parameter $ A_{\max} $ and  base frequency $ f_0 $ are as given in \textbf{Table \ref{tab:sim-data}}}\label{fig:std-vs-nit}
\end{figure}

\begin{figure}
	\centering
	\includegraphics[scale=0.85]{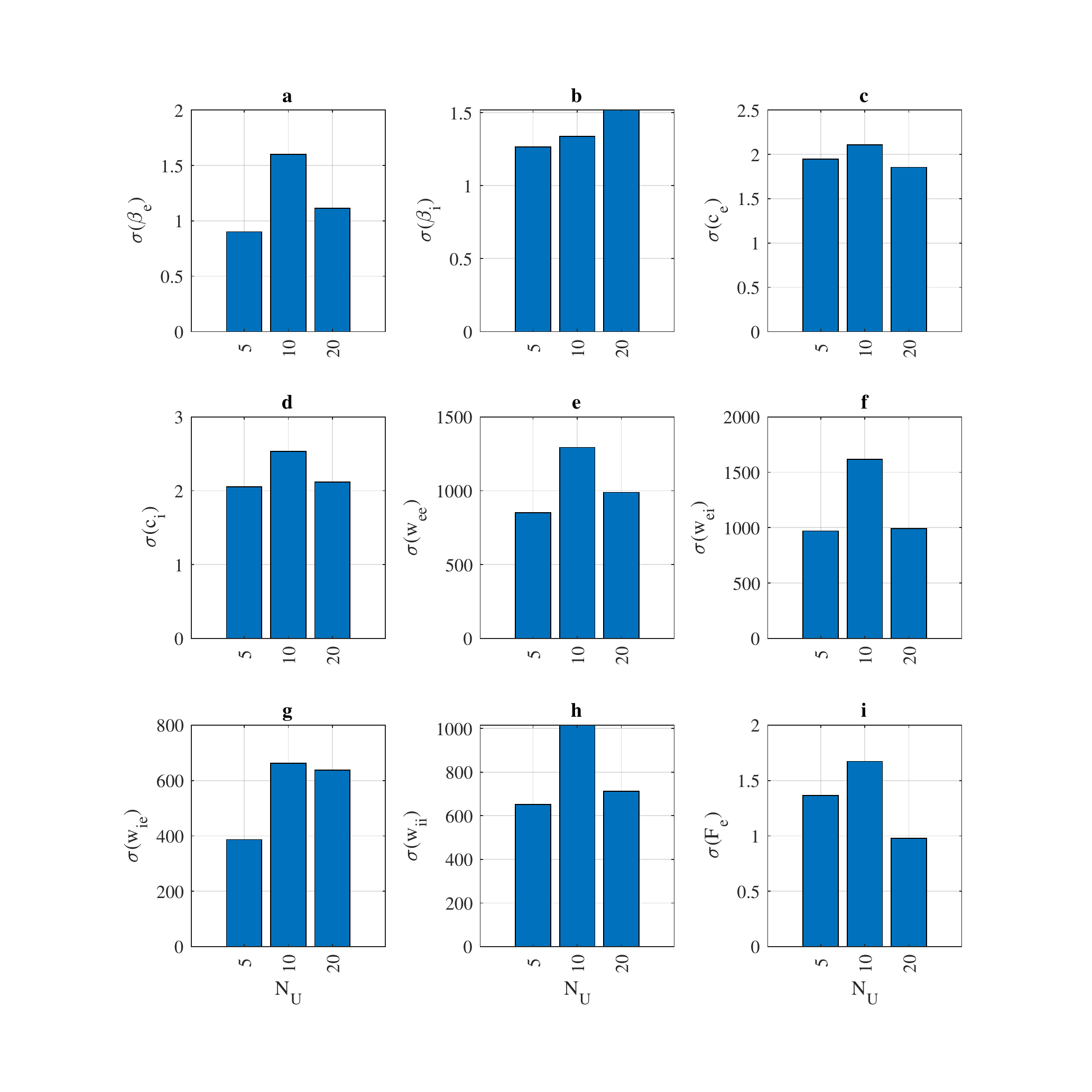}
	\caption{The variation of the standard deviations of the estimated parameters against stimulus component size $ N_{U} $. The settings of sample size $ N_{it} $, maximum amplitude parameter $ A_{\max} $ and  base frequency $ f_0 $ are as given in \textbf{Table \ref{tab:sim-data}}}\label{fig:std-vs-nu}
\end{figure}

\begin{figure}
	\centering
	\includegraphics[scale=0.85]{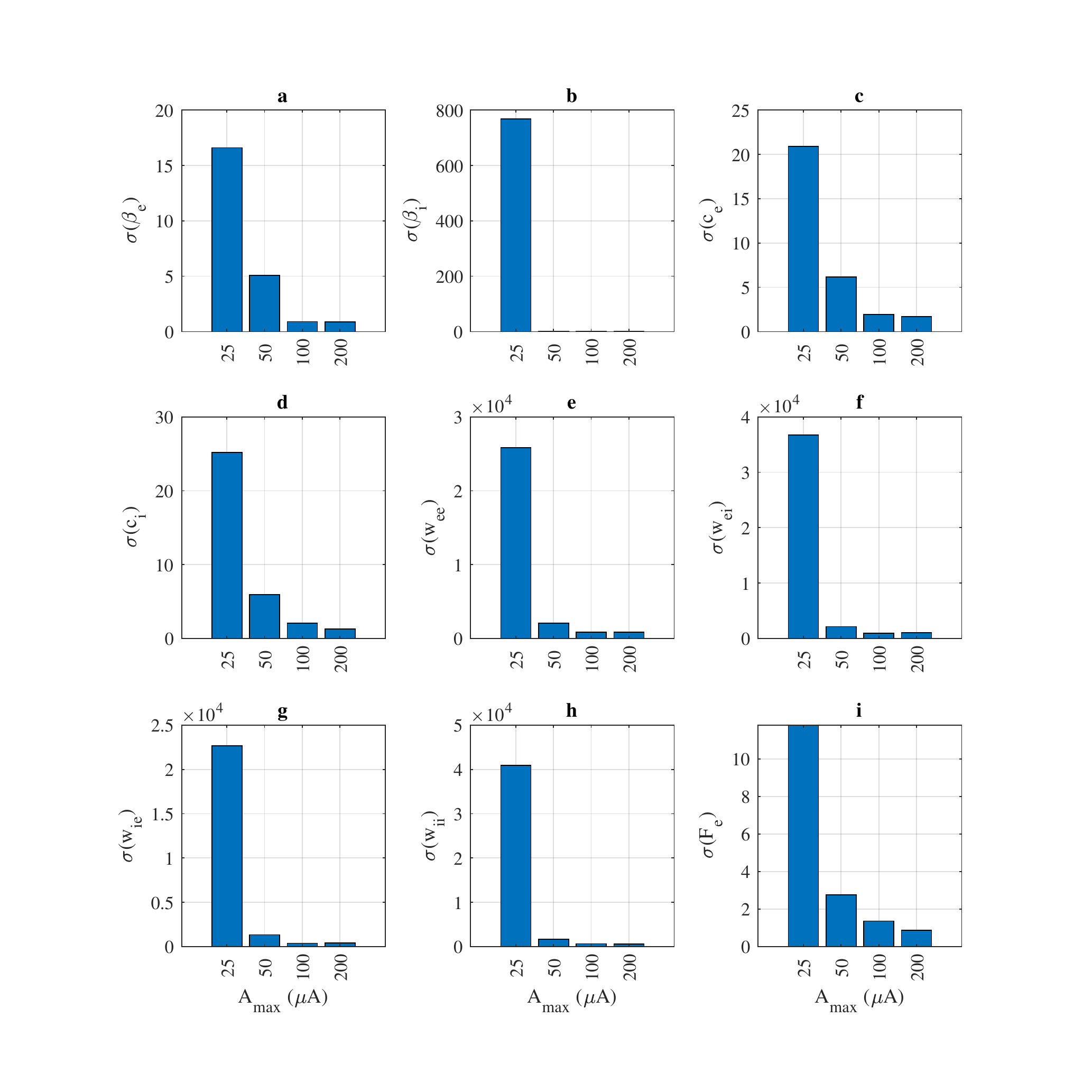}
	\caption{The variation of the standard deviations of the estimated parameters against maximum amplitude parameter $ A_{\max} $. The settings of sample size $ N_{it} $, stimulus component size $ N_{U} $ and  base frequency $ f_0 $ are as given in \textbf{Table \ref{tab:sim-data}}}\label{fig:std-vs-amax}
\end{figure}

\begin{figure}
	\centering
	\includegraphics[scale=0.85]{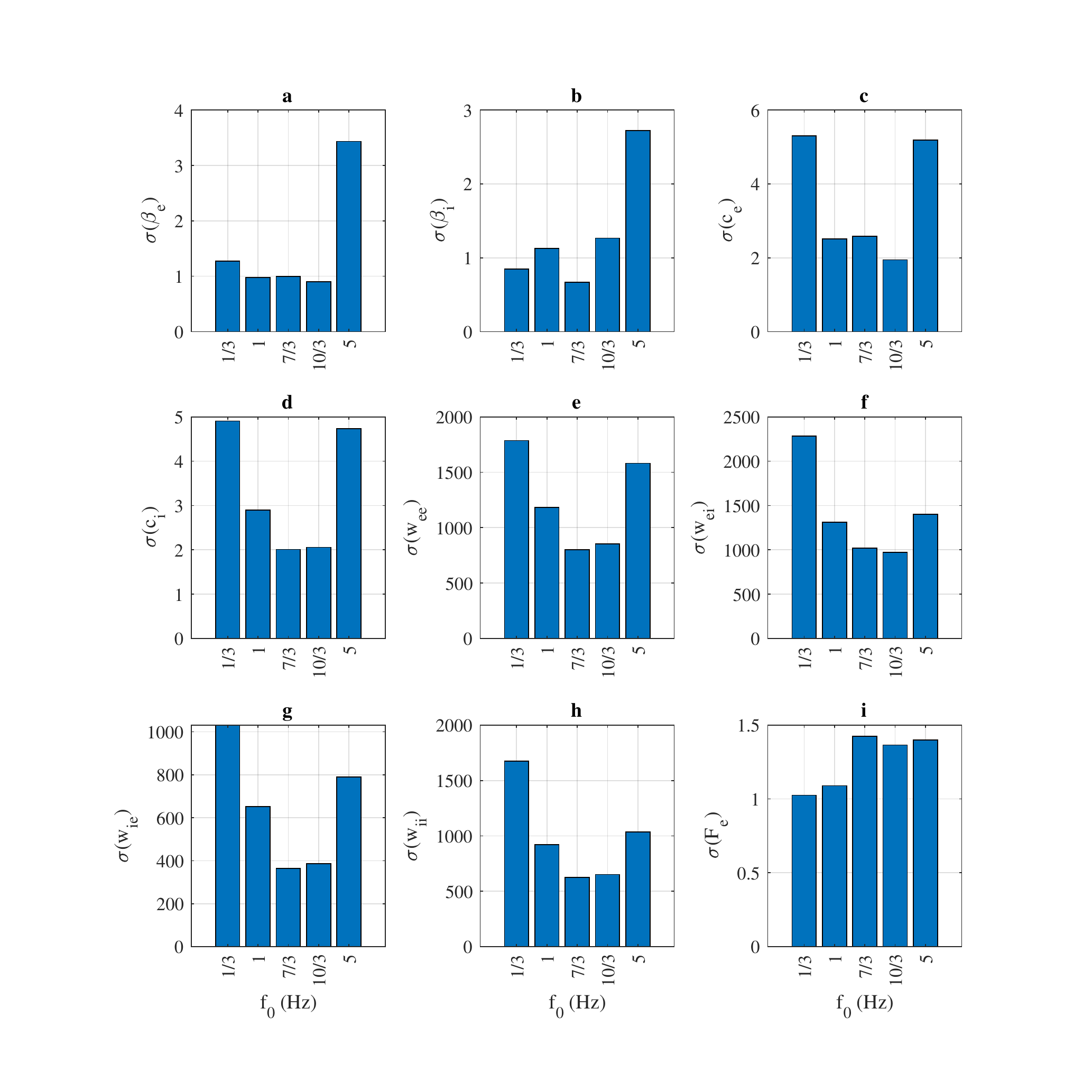}
	\caption{The variation of the standard deviations of the estimated parameters against base frequency $ f_0 $. The settings of sample size $ N_{it} $, stimulus component size $ N_{U} $ and  maximum amplitude parameter $ A_{\max} $ are as given in \textbf{Table \ref{tab:sim-data}}}\label{fig:std-vs-f0}
\end{figure}

One can see the variation of the standard deviation of the estimates against the sample size $ N_{it} $, stimulus component size $ N_U $, maximum amplitude parameter $ A_{\max} $ and stimulus base frequency $ f_0 $ in \textbf{Figures \ref{fig:std-vs-nit}}, \textbf{\ref{fig:std-vs-nu}}, \textbf{\ref{fig:std-vs-amax}}  and \textbf{\ref{fig:std-vs-f0}} respectively. 

\subsection{Comparing the Outputs of Original and Estimated Models}
\label{sec:mse-test}
\begin{figure}
	\centering
	\includegraphics[scale=0.85]{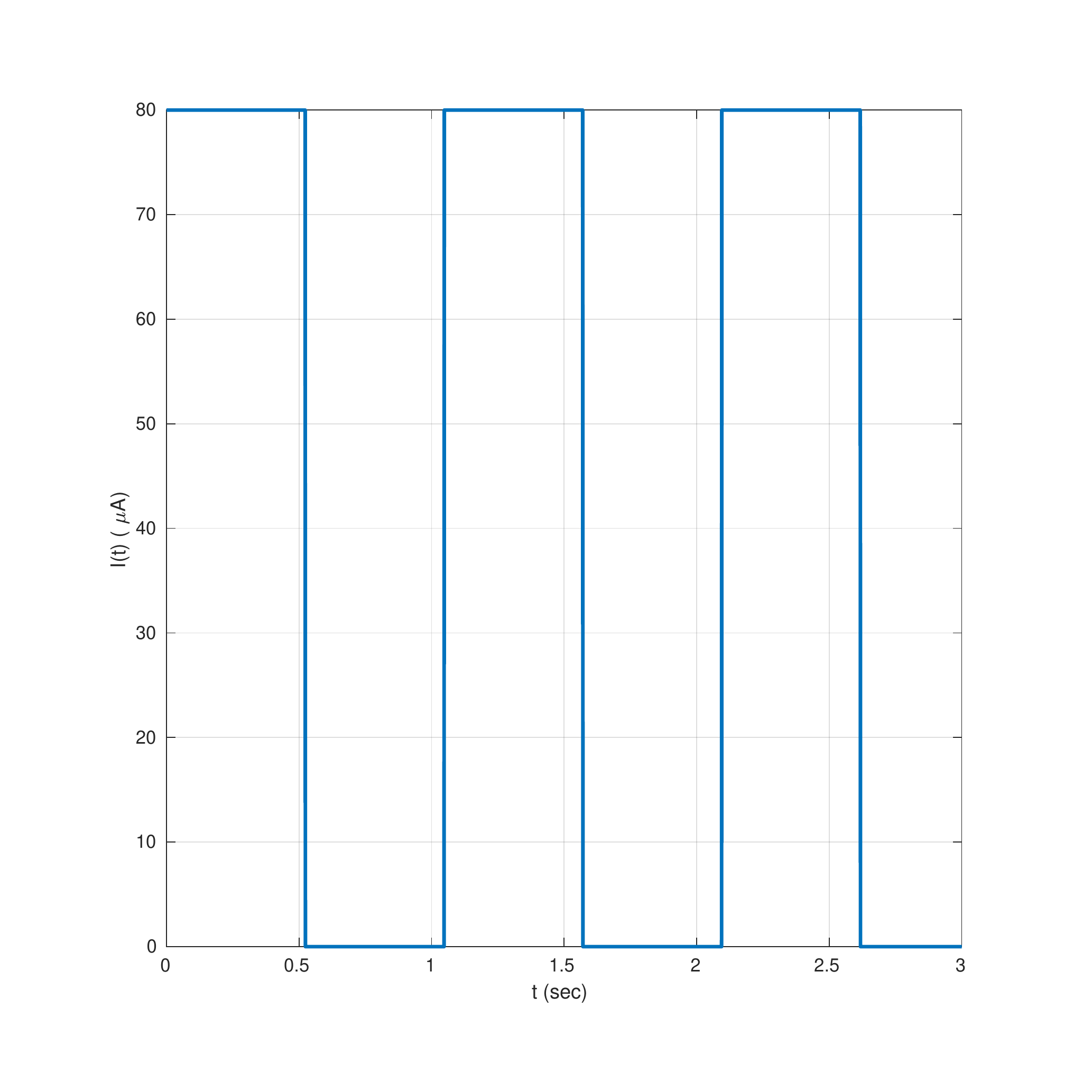}
	\caption{Square wave test stimulus}\label{fig:test-stim-sq}
\end{figure}

\begin{figure}
	\centering
	\includegraphics[scale=0.85]{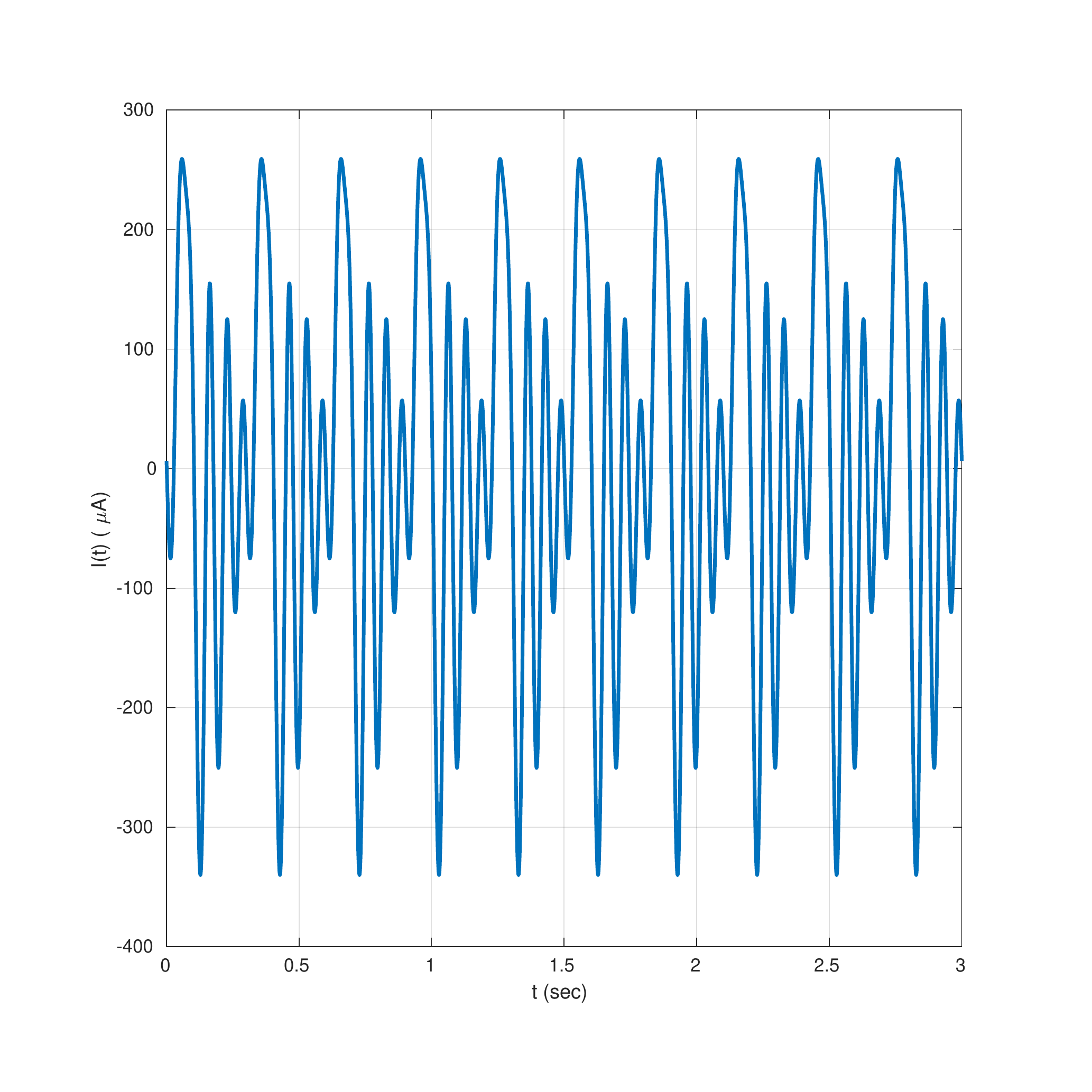}
	\caption{Fourier series test stimulus}\label{fig:test-stim-fs}
\end{figure}

In order to validate the success of our work, we will need to check whether the firing rate outputs of two models are tracking each other. This can be performed by applying a stimulus profile to both original or data generator model in \eqref{eq:data-gen-ctrnn-model} and our estimated model in \eqref{eq:gen-ctrnn} and compare their firing rate outputs. It will be better to test the models using at least two different stimuli. In this research, we will do this by driving our models by a square wave stimulus such as the one shown in \textbf{Figure \ref{fig:test-stim-sq}} and a Fourier series one such as the one shown in \textbf{Figure \ref{fig:test-stim-fs}}. 

\begin{figure}
	\centering
	\includegraphics[scale=0.85]{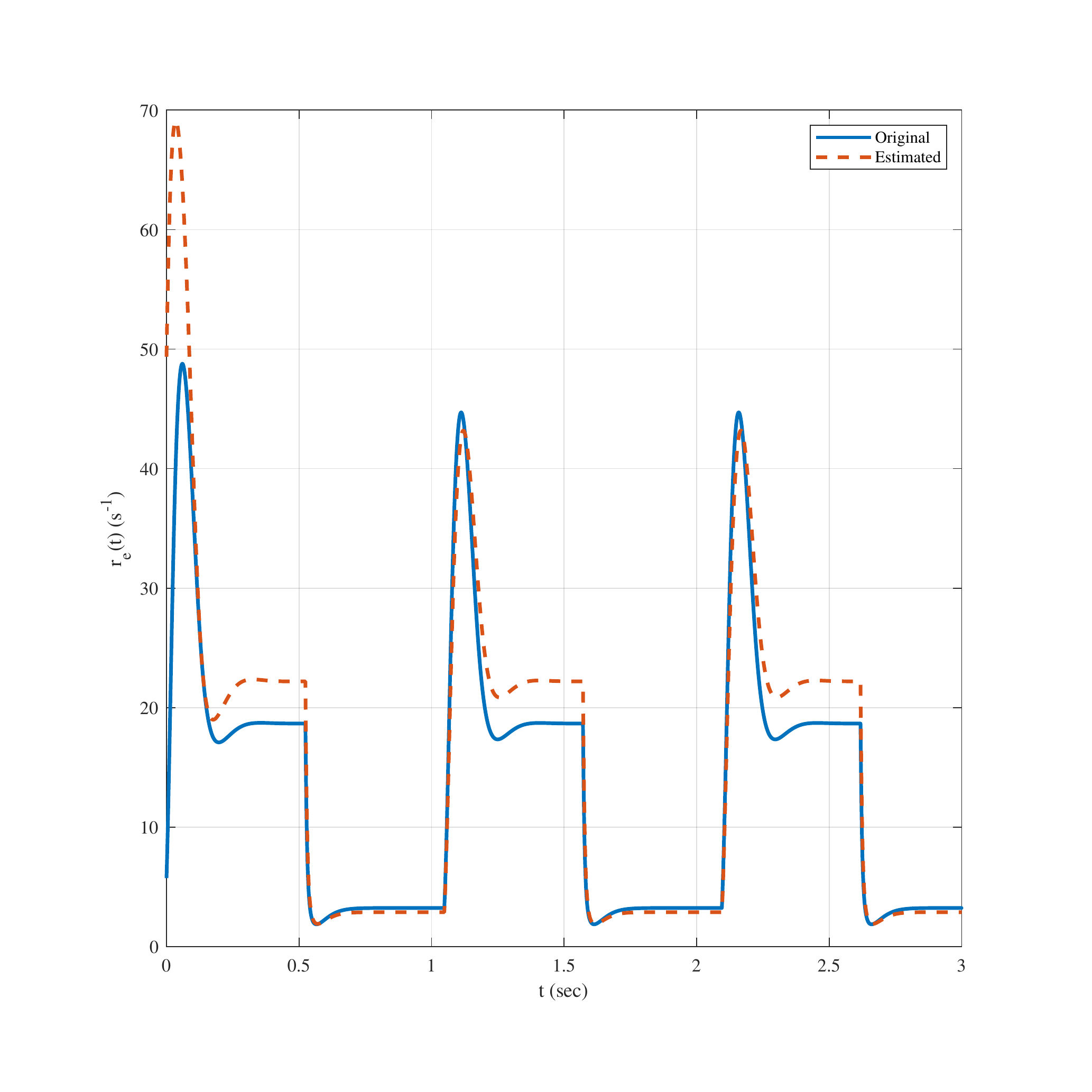}
	\caption{Response of \eqref{eq:gen-ctrnn} to square wave stimulus in \textbf{Figure \ref{fig:test-stim-sq}}. The model parameters are taken from the last row of \textbf{Table \ref{tab:mean-vals-vs-nit}}. }\label{fig:test-resp-sq}
\end{figure}

\begin{figure}
	\centering
	\includegraphics[scale=0.85]{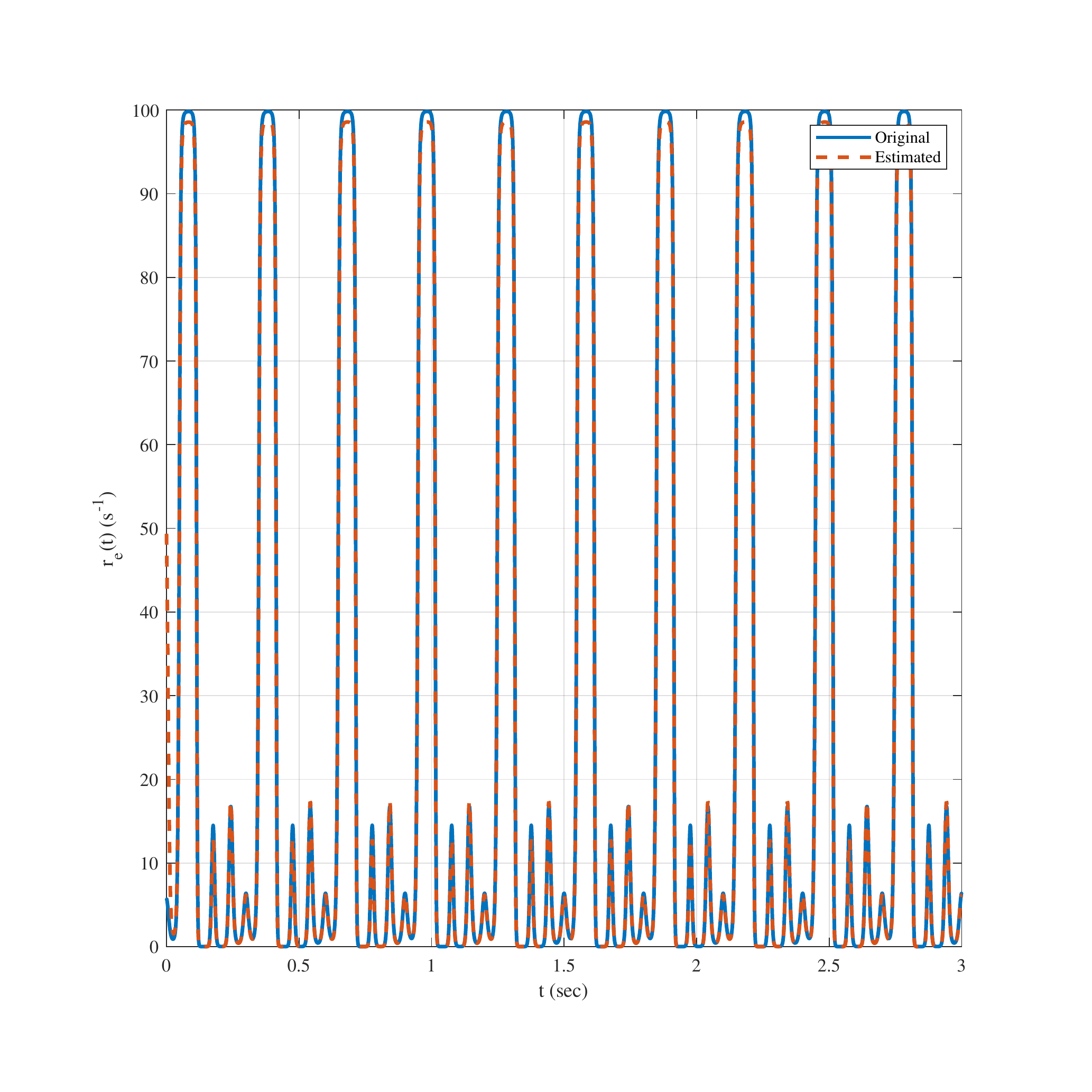}
	\caption{Response of \eqref{eq:gen-ctrnn} to square wave stimulus in \textbf{Figure \ref{fig:test-stim-fs}}. The model parameters are taken from the last row of \textbf{Table \ref{tab:mean-vals-vs-nit}}.}\label{fig:test-resp-fs}
\end{figure}

The response obtained from \eqref{eq:gen-ctrnn} with the estimated parameters corresponding to the case $ N_{it}=400 $, $ N_U=5 $, $ A_{\max}=100 $ and $ f_0=3.3333 $ (the last case in \textbf{Table \ref{tab:mean-vals-vs-nit}}) to the stimuli in \textbf{Figures \ref{fig:test-stim-sq}} and \textbf{\ref{fig:test-stim-fs}} can be seen in \textbf{Figures \ref{fig:test-resp-sq}} and \textbf{\ref{fig:test-resp-fs}} respectively. 

\section{Discussion}

\subsection{Summary}

In this research, we considered a theoretical study of model fitting to a noisy stimulus/response data obtained from a generic sensory neural network. In general the purpose is similar to that that of \cite{doruk2018}. The main differences can be summarized as shown below:

\begin{enumerate}
	\item We desire a more generic model that relates stimulus to firing rate response of the neural system under consideration. Such a model has smaller number of parameters. 
	\item In \cite{10.7287/peerj.preprints.27015v1}, it is found that inclusion of sigmoidal gain function parameters to the estimation problem leads to a degradation in estimation performance. For example, increasing the sample size $ N_{it} $, did not improve the variance of estimates.  
\end{enumerate}

Thus, we preferred a model which does not require the estimation of the sigmoidal gain function parameters. That will decrease the level of parameter confounding so that we expect a reduced level of estimation errors. In addition data is taken from a data generator model which is different in form (the model used in \cite{doruk2018}). This means that we do not have any true parameters and thus we will need to compare the firing rate responses of the data generator (original model in other words) and the estimated model. Upon completion of the study we noticed the results presented in the next section. 

\subsection{Discussion of the Results}

In \textbf{Tables \ref{tab:mean-vals-vs-nit}}, \textbf{\ref{tab:mean-vals-vs-nu}}, \textbf{\ref{tab:mean-vals-vs-amax}} and \textbf{\ref{tab:mean-vals-vs-f0}} one can see the mean estimated values of the parameters of our model in \eqref{eq:gen-ctrnn} and \eqref{eq:firing-rate-sigmoid}. The values are remaining in a narrow range against changing simulation parameters sample size $ N_{it} $, stimulus component count $ N_U $, maximum amplitude $ A_{\max} $ and base frequency $ f_0 $ parameters. One caveat appears when $ A_{\max}=25 $, the parameter $ w_{ei} $ and $ w_{ie} $ deviates from their values correspoding to the amplitudes larger than $ A_{\max}=25 $. This result verifies that, the nominal selection of $ A_{\max}=100 $, $ N_U=5 $ and $ f_0=10/3 $ is a good compromise. 

The results associated with the variation of standard deviations against increasing simulation parameters are available in \textbf{Figures \ref{fig:std-vs-nit}}, \textbf{\ref{fig:std-vs-nu}}, \textbf{\ref{fig:std-vs-amax}} and \textbf{\ref{fig:std-vs-f0}}. 

These results are as expected. The standard deviations of estimates of each parameter decreases with increasing sample size $ N_{it} $. This occurrence is similar to that of \cite{doruk2018}. 

There isn't any definite pattern associated with the changing stimulus component size $ N_U $. Thus, there is no need to choose large $ N_U $ values that brings unnecessary computational complexity.    

Concerning the stimulus amplitude $ A_{\max} $, one can say that lower levels should be avoided. \textbf{Figure \ref{fig:std-vs-amax}} suggests that, choosing a value like $ A_{\max}=25 $ will negatively affect the estimation performance. Same graphical results suggest that a selection of $ 100 \leq A_{\max} \leq 200 $ should be a good compromise. 

The changes in base frequency is also effective on the estimate variance. \textbf{Figure \ref{fig:std-vs-f0}} suggests that mid-range frequencies ($ \frac{7}{3} \leq f_0 \leq \frac{10}{3} $ Hz) should be preferred. 

Since we are estimating a model of which true parameters are not known, we needed to develop a verification mechanism. This can be done by comparing the firing rate responses of the original or data generator model and the estimated model. In this research, we did that by stimulating both models by a square wave (\textbf{Figure \ref{fig:test-stim-sq}}) and a Fourier series (\textbf{Figure \ref{fig:test-stim-fs}}) stimuli. The corresponding responses were given in \textbf{Figure \ref{fig:test-resp-sq}} against square wave and \textbf{Figure \ref{fig:test-resp-fs}} against Fourier series stimuli respectively. One can say that under a smooth stimulus like Fourier series both original (data generator) and estimated models generate almost same firing rate response (\textbf{Figure \ref{fig:test-resp-fs}}). When a discontinuous stimulus like \textbf{Figure \ref{fig:test-stim-sq}} is applied, the responses elicited by both models have a small deviation at the discontinuities \textbf{Figure \ref{fig:test-resp-fs}}. 

Based on all those findings one can state that, the estimation procedure is completed successfully.

\section*{Acknowledgement}
The computational facilities needed by this research are provided by TR-GRID/TRUBA framework ({www.truba.gov.tr}) operated by National Academic Center of Computing (ULAKBIM) of Turkish Scientific and Technological Research Institution (TUBITAK).



\end{document}